# Heat transfer enhancement of N-Ga-Al semiconductors heterogeneous interfaces

Wenzhu Luo [a†], Ershuai Yin [a†*], Lei Wang [a], Wenlei Lian [bc], Neng Wang [a], Qiang Li [a*]

[a] MIIT Key Laboratory of Thermal Control of Electronic Equipment, School of Energy and Power Engineering, Nanjing University of Science & Technology, Nanjing, Jiangsu 210094, China

[b] College of Energy and Power Engineering, Nanjing University of Aeronautics and Astronautics, Nanjing, Jiangsu 210016, China

[c] Key Laboratory of Thermal Management and Energy Utilization of Aviation Vehicles, Ministry of Industry and Information Technology, Nanjing, Jiangsu 210016, China

**Abstract:** Heat transfer enhancement of N-Ga-Al semiconductor heterostructure interfaces is critical for the heat dissipation in GaN-based electronic devices, while the effect of the $Al_xGa_{(1-x)}N$ transition layer component concentration and thickness on the heat transfer mechanism at the GaN-AlN interface is unclear. In this paper, using molecular dynamics simulations based on machine learning potentials, the interfacial thermal conductance (ITC) between $GaN-Al_xGa_{(1-x)}N$, $AlN-Al_xGa_{(1-x)}N$ and $GaN-Al_xGa_{(1-x)}N-AlN$ heterostructure interfaces are calculated for different transition layer thicknesses with different concentrations of Al fractions, and the reasons for the change of ITC and its heat transfer mechanism were explained by the phonon density of states and the spectral heat current. GaN-AlN heterostructure ITC at 300 K is calculated to be 557 $MW/(m^2K)$, and the ITCs of $GaN-Al_{0.5}Ga_{0.5}N$ and $AlN-Al_{0.5}Ga_{0.5}N$ are improved by 128% and 229% compared to GaN-AlN, whereas the ITCs of $GaN-Al_{0.7}Ga_{0.3}N-AlN$ containing a 0.5 nm transition layer improved by 27.6%. This is because elemental doping enhances phonon scattering near the interface thereby promoting phonon energy redistribution, but the bulk thermal resistance of the $Al_xGa_{(1-x)}N$ layer also increases rapidly with increasing doping ratio, and ITC is affected by a combination of these two factors. This work aims to understand the mechanism of transition layer component concentration and thickness on the heat transfer at the GaN-AlN contact interface, which provides a useful guide for better thermal design of the GaN-AlN heterostructure interface.

**Keywords:** GaN-AlN heterostructure interface, interfacial heat transfer, machine learning interatomic potentials, molecular dynamics

---

[†] These authors contributed equally to this work.

[*] Corresponding authors. E-mail address: yes@njust.edu.cn (E. Yin), liqiang@njust.edu.cn (Q. Li)

# 1 Introduction

III-V compound semiconductor material Gallium nitride (GaN) has a wide band gap [1], high electron mobility [2], and high electron saturation velocity [3], GaN-based high electron mobility transistors (HEMTs) have a wide range of prospects for high-frequency, high-power electronics applications [4,5], such as RF amplifiers [6], power amplifiers for satellite base stations and radar sensors [7,8]. The unavoidable Joule self-heating effect [9] within the two-dimensional electron gas channel can greatly affect the device lifetime and efficiency [10], while the thermal boundary resistance (TBR = 1/TBC) between the channel layer and the substrate material accounts for a large portion of the thermal resistance of the near junction, which seriously hampers the heat transfer to become a bottleneck in the development of GaN-based electronic devices [11]. Therefore, conducting GaN-based semiconductor hetero-interface heat transfer studies to reduce the hot spot temperature is crucial to enhance device performance and maintain stability. Currently, studies report the enhancement of interfacial heat transfer through transition layers, nanostructures [12,13] and improvement of interfacial bond strength [14].

Due to the lattice mismatch between GaN and substrate, Aluminum nitride (AlN) is commonly used as a transition layer material between GaN/substrate heterostructure interface. On the experimental side, Zhou et al [15] found that the AlN transition layer can effectively reduce the effective thermal boundary resistance ($TBR_{eff}$) at the GaN/Diamond interface; Li et al [16] explored the effect of different AlN layer thicknesses on the GaN/SiC thermal boundary conductance (TBC); Cho et al [17] measured the temperature dependence of TBR of GaN/AlN/Si heterostructures in the 300- to 550 K the temperature dependence of TBR. Due to the large lattice mismatch between Si and GaN, stresses are also usually released through the structure of AlN nucleation layers and $Al_xGa_{(1-x)}N$ gradient buffer layers [18,19], and it has been experimentally found that atomic diffusion at the interface of GaN-AlN heterostructures has always existed [20,21]. Therefore, the GaN-$Al_xGa_{(1-x)}$N-AlN heterostructure interface is a common structure in GaN-based devices, which is important to study the mechanism of the influence of transition layers with different Al component concentrations and thicknesses on the interfacial heat transport in this heterostructure.

The study of interfacial heat transport properties requires accurate and efficient computational methods, and molecular dynamics (MD) can effectively simulate the effect of all physical phenomena near the interface of a real physical model on interfacial heat transport [22], and the accuracy of its prediction depends on the precision of the interatomic potential function [23]. Density Functional Theory (DFT) approximations are computed with high accuracy but require extensive quantum calculations to obtain the phonon thermal properties [24]. Recently developed machine learning interatomic potentials(MLIPs) can be well combined with first-principles calculations and molecular dynamics to efficiently and accurately probe the heat transport mechanism at heterogeneous interfaces, and for this reason many MLIPs models have been developed, such as the Neuroevolution-potential (NEP)[25], Neural network potential(NNP) [26], and Gaussian approximation potential(GAP) [27]. Sun et al [28] investigated the effect of crystal orientation on the thermal boundary resistance of the β-$Ga_2O_3$/Diamond heterostructure interface using a constructed neural evolution potential(NEP); Cheng et al [29] trained a high-fidelity neural network potential(NNP) for MD simulations at the Si/Ge interface, and the predicted thermal boundary conductance(TBC) was in good agreement with the experimental measurements; Huang et al [30] calculated the interfacial thermal conductance of GaN/AlN heterostructures at different interface morphologies using a deep neural network potential(NNP), but did not explore the effect of the interlayer on interfacial heat transfer in depth. However, Fan et al [31] used a separable natural evolution strategy to train a neural network-based machine learning potential NEP, which demonstrated superior accuracy and computational efficiency compared to other MLIPs by using a highly efficient Graphics Processing Unit (GPU) for computation in the GPUMD software package [32].

In this work, we constructed the NEP of GaN-AlN heterostructure interface and used nonequilibrium molecular dynamics (NEMD) to calculate the interfacial thermal conductance (ITC) of GaN-$Al_xGa_{(1-x)}$N, AlN-$Al_xGa_{(1-x)}$N, and GaN-$Al_xGa_{(1-x)}$N-AlN, which contain $Al_xGa_{(1-x)}$N transition layers of different component concentrations and thicknesses. Furthermore, phonon density of states (PDOS) and spectral heat current (SHC) are calculated to reveal the effects of transition layer materials with different component concentrations and thicknesses on the heat transfer at the GaN-AlN interface and to investigate the interfacial heat transport mechanism.

## 2 Computational methodology

### 2.1 Brief introduction to the NEP model

Using a separable natural evolution strategy implemented in the GPUMD software package to train the neural network-based machine learning potential NEP. The site energy of atom $i$ is taken as a function of the descriptor vector $U_i$ with $N_{des}$ components and this function is modeled by a feedforward neural network with $N_{neu}$ neurons [33]:

$$U_i = \sum_{\mu=1}^{N_{neu}} w_\mu^{(1)} \tanh\left(\sum_{v=1}^{N_{des}} w_{\mu v}^{(0)} q_v^i - b_\mu^{(0)}\right) - b^{(1)} \tag{1}$$

Where tanh($x$) is the activation function in the hidden layer, $w^{(0)}$ is the connection weight matrix from the input layer to the hidden layer, $b^{(0)}$ is the bias vector in the hidden layer, $w^{(1)}$ is the connection weight vector from the hidden layer to the output node, $b^{(1)}$ is the bias of the output layer node $U_i$, $N_{des}$ is the number of components of the descriptor vector and $N_{neu}$ is the number of neurons. A simplified schematic of the NEP model framework is shown in Fig 1(a), the hidden layer of this study has 50 neurons. More detailed derivations and theoretical formulations of the NEP model can be found in reference [34].

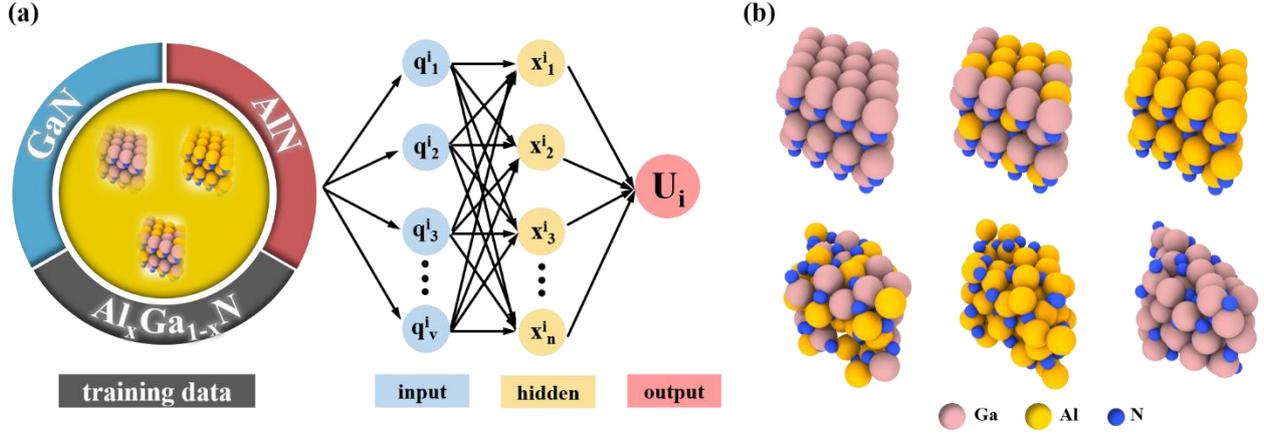

**Fig. 1.** (a) Simplified schematic of the NEP framework. (b) Partial NEP training dataset structural model.

## *2.2 Training dataset and NEP training*

To investigate the mechanism of the transition layer material and thickness on the heat transfer at the GaN-AlN heterostructure interface, a training dataset containing three elements of Ga-N-Al was constructed. For the crystal structure, the GaN with wurtzite structure was first expanded to obtain a supercell containing 128 atoms. Since the investigated structures contain complex $Al_xGa_{(1-x)}N$ hybrid systems, the training dataset needs to include a large number of relevant sample structures, where Ga atoms in GaN supercells are randomly replaced with Al atoms in the ratio of 0.1 to 1, with an interval of 0.1. Therefore, a total of 11 original crystal structures containing 1 GaN, 9 $Al_xGa_{(1-x)}N$ (x = 0.1-0.9), and 1 AlN were obtained, followed by melt annealing, random perturbation, relaxation and compression of the 11 original structures to generate 11 amorphous, 22 perturbed and 22 scaled structures. Firstly, the original 11 structures were heated at 7000 K for 1 ps, and after the structures were completely melted they were cooled to room temperature within 1.5 ps to obtain disordered amorphous structures. In addition, each original structure was relaxed and compressed to obtain scaled structures with scaling factors of 0.97 and 1.03. The perturbed structure is obtained by randomly perturbing the atomic positions in the original structure, where the perturbation ratio is 0.05 and the maximum perturbation distance is 0.1 Å. After obtaining 11 original and 55 extended structures, AIMD calculations were performed on all the structures using the Vienna Ab initio Simulation Package (VASP) at 200-600 K to obtain the total energies, interatomic forces, and virials information for the corresponding configurations. Finally, 4400 original structures, 4400 amorphous structures, 2200 scaled structures, and 2200 perturbed structures were extracted, for a total dataset of 13,200, which was randomly divided into training and test datasets in a ratio of 9:1. A model of the structure of the partial NEP training dataset is shown in Fig. 1(b), listing the original structure and amorphous structure of GaN, AlN and $Al_xGa_{(1-x)}N$. Reasonable hyperparameters and appropriate descriptors are the keys to obtaining high-accuracy NEP, the training hyperparameters in this paper are shown in Table 1.

The total loss function, energy, and force loss terms of the training and test datasets show very good convergence after 800000 generations of iterations, as shown in Fig. 2(a). The accuracy of the NEP model was verified by comparing the NEP predicted and DFT calculated energies and forces, as shown in Fig. 2(b-c), the results of the NEP predictions and DFT calculations were uniformly distributed around the diagonal line showing good agreement, and the root-mean-square error(RMSE) for the training dataset energies and forces were 4.34 meV/atom and 125 meV/Å, and the test dataset energies and force have RMSE values of 3.17meV/atom and 86.58meV/Å respectively. Usually trained MLIPs with energy terms in the range of several meV/atom and force terms in the range of several hundred meV/Å are trained satisfactorily [35], and the small RMSE values also prove the high quality of the NEP models we constructed. Phonon dispersion as the most important material thermal property is also often used to evaluate the interatomic potential [36], as shown in Fig. 2(d-e), we calculated the phonon dispersion of GaN and AlN by using the fitted NEP model, and the predicted results are in good agreement with the experimental data [37,38], which proves that the constructed NEP can accurately predict the material thermophysical properties, and thus it can be used to explore the GaN-AlN heterostructure interface heat transfer mechanism.

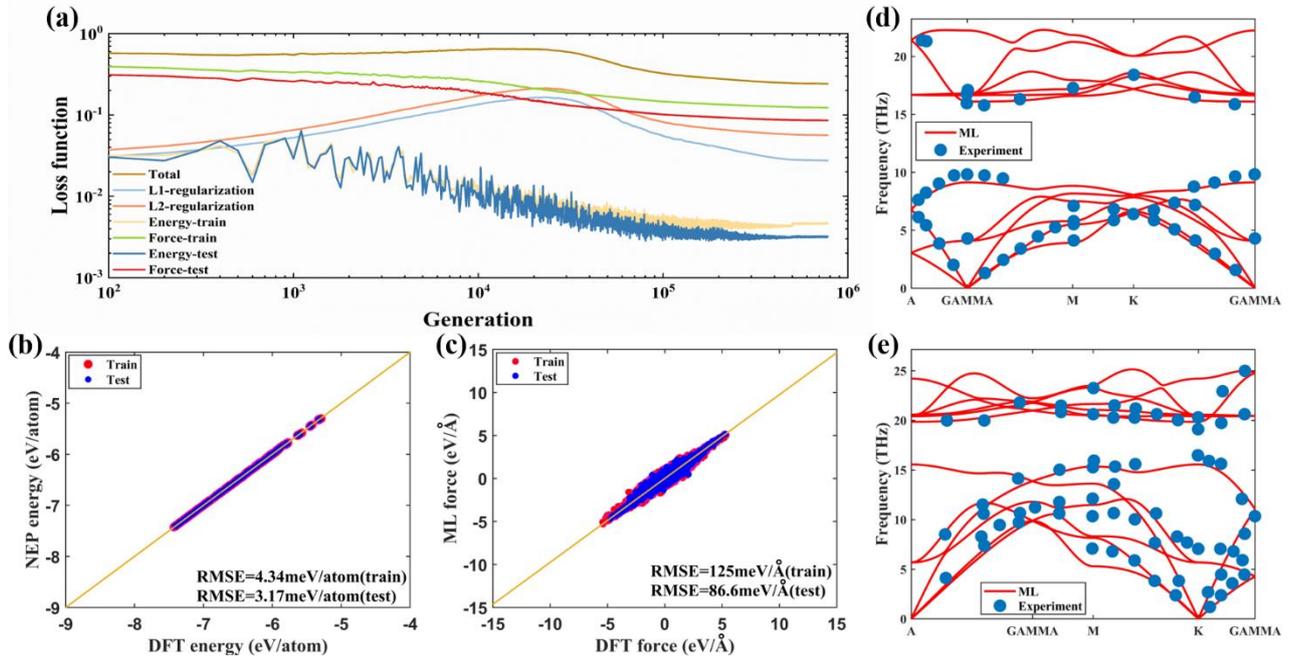

**Fig. 2.** (a) Evolution of the loss function during training, L1 and L2 are regularisation coefficients. Comparison between NEP predictions and DFT calculated values for (b) energy and (c) force for the training and test sets. NEP predictions versus experimentally measured (d) GaN and (e) AlN phonon dispersion.

**Table 1.** NEP training hyperparameters

| parameter | value | parameter | value |
| --- | --- | --- | --- |
| $r_C^R$ | 6 | $r_C^A$ | 4 |
| $n_{max}^R$ | 8 | $n_{max}^A$ | 8 |
| $N_{basis}^R$ | 12 | $N_{basis}^A$ | 10 |
| $l_{max}^{3b}$ | 4 | $l_{max}^{4b}$ | 2 |
| $N_{neuron}$ | 50 | $N_{population}$ | 80 |
| $N_{batch}$ | 3000 | $N_{generation}$ | 800000 |
| $\lambda_1$ | 0.1 | $\lambda_2$ | 0.1 |
| $\lambda_e$ | 0.5 | $\lambda_f$ | 1.2 |
| $\lambda_v$ | 0.1 | type | N Ga Al |

*2.3 Non-equilibrium molecular dynamics simulation*

Nonequilibrium molecular dynamics (NEMD) simulations were used to calculate the interfacial thermal conductance (ITC) for all models, thus investigating the mechanism of the transition layer material and thickness on the heat transfer at the GaN-AlN heterostructure interface. The ITC is calculated using the heat flux $J$ through the interface and the temperature drop $\Delta T$ at the interface, and the ITC is defined as:

$$ITC = \frac{J}{\Delta T} \qquad (2)$$

The GaN and AlN expanded cells were combined into a smooth heterostructure interface structure as our base model, and then different ratios of Ga and Al atoms in the base model are randomly replaced to generate different concentrations of $Al_xGa_{(1-x)}N$ layers. Four heterostructures of GaN-AlN, GaN-$Al_xGa_{(1-x)}N$, $Al_xGa_{(1-x)}N$-AlN, and GaN-$Al_xGa_{(1-x)}N$-AlN were finally constructed. The computational model uses GaN and AlN with wurtzite structure, with lattice constants $a_{GaN}$ = 3.176 Å, $c_{GaN}$ = 5.176 Å, $a_{AlN}$ = 3.130 Å, and $c_{AlN}$ = 5.011 Å, respectively. The GaN and AlN expand their cells by 9, 5, and 20 in the x, y, and z directions, respectively, and the total dimensions of the simulated box are W × H × L = 28.17 × 27.11 × 205.75 Å, where W, H and L are the width, height and length of the box, the GaN and AlN contact interface mismatch rate is 4.6%, and the total number of atoms in the model is 14400. The GaN surface is terminated by Ga atoms at the GaN-AlN interface, which is consistent with the actual interfacial growth of GaN and AlN.

Considering the practical application of GaN-based devices, a heat source (T = 330 K) and a heat sink (T = 270 K) are applied at the end of GaN and AlN, respectively, and the heat flux flows from GaN to AlN, as shown in Fig. 3. Periodic boundary conditions are imposed in the x and y directions, and fixed boundary conditions are set in the z-direction, dividing the simulation box into 42 slices in the z direction, and fixed for atoms in the outermost 4.9 Å range on each side of the z-direction. The time step was set to 1fs, firstly, the model was relaxed under the canonical ensemble (NVT) for 1ns to steady the temperature at 300K, then a Langevin thermostat was used to stabilize the heat flux under the NVE ensemble for 1ns, and finally 0.5ns was computed under the same conditions for the data outputs. The relevant data was sampled once every 10 steps averaged 5000 times to output the results once, and the final temperature distribution was obtained by averaging the 10 data calculated for the last 0.5ns. In this paper, NEP training and molecular dynamics simulations were performed using the open-source GPUMD software package.

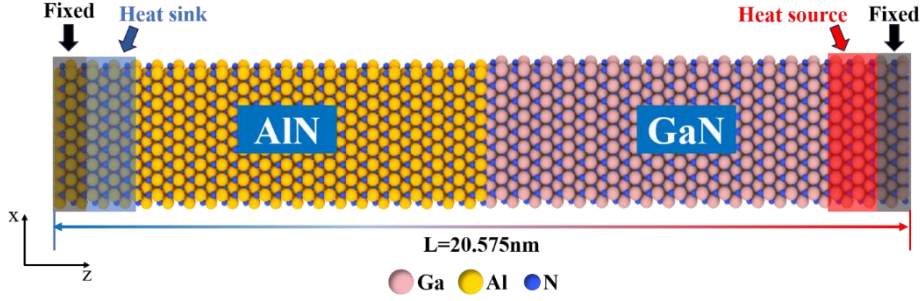

**Fig. 3.** Schematic of NEMD simulation of GaN-AlN heterostructure.

*2.4 The spectral heat current (SHC)*

The spectral decomposition of nonequilibrium heat current[39,40] can obtain the contribution of different frequency phonons to the interfacial heat flux, and the calculation of spectral heat current (SHC)[41] and spectral thermal conductivity [42] by this method has been widely used to explore the interfacial heat transfer mechanism. Firstly, the following steady-state time-correlation function is defined:

$$K(t) = \sum_i \sum_{j \neq i} r_{ij}(0) \left\langle \left( \frac{\partial U_j}{\partial r_{ji}}(0) \cdot \frac{p_i(t)}{m_i} \right) \right\rangle \quad (3)$$

Where $r_i$, $m_i$, and $p_i$ are the position, mass, and momentum of particle $i$, $r_{ij}$ is defined as $r_i$-$r_j$. It reduces to a nonequilibrium heat current when t = 0. One can define the following Fourier transforms:

$$\tilde{K}(\omega) = \int_{-\infty}^{\infty} dt e^{i\omega t} K(t), K(t) = \int_{-\infty}^{\infty} \frac{d\omega}{2\pi} e^{-i\omega t} \tilde{K}(\omega) \quad (4)$$

For the second equation above, setting t = 0 yields the following spectral heat current (SHC):

$$J_q(\omega) = 2\tilde{K}(\omega), \left\langle J_q \right\rangle_{ne} = \int_0^{\infty} \frac{d\omega}{2\pi} J_q(\omega) \quad (5)$$

The computational model and the relaxation process for SHC are the same as in the NEMD setup, considering the spectral decomposition of the heat current in the z-direction with a maximum angular frequency of 400 THz. Detailed formula derivation and application of SHC can be found in reference [43].

## 3 Results and discussion

*3.1 GaN-AlN interfacial heat transfer*

The interfacial heat transfer of the GaN-AlN heterostructure interface at 300 K was investigated by NEP-based NEMD simulations. As shown in Fig. 4(a) for the temperature distribution when the GaN-AlN heterostructure is stabilized, it can be found that there is an obvious temperature jump ($\Delta T$) at the interface, indicating that there is a finite interfacial thermal resistance (ITR) between GaN and AlN. Fig. 4(b) shows the energy values of the inflow heat source and outflow heat sink of the system during the NEMD simulation, which can be found to be highly symmetric concerning E = 0 to satisfy the law of energy conservation. In addition, since the results of NEMD calculations may be affected by the size effect, we performed the calculations with systems of different sizes to eliminate the size effect, as shown in Fig. 4(c) for the effect of different model lengths on ITC, the interfacial cross-sectional area is fixed at $28.17 \times 27.11$ Å, and it can be found that the difference in interfacial thermal conductance under different model lengths is within 6%, when the length is 20 nm the ITC has been stabilized, so the model size for all of our subsequent simulations is $28.17 \times 27.11 \times 205.75$ Å.

The NEP model predicts an ITC of 557 MW/(m$^2$K) for GaN-AlN heterostructures at 300 K, which is very close to the experimental results (ITC = 620 MW/(m$^2$K)) measured by Koh et al [44] via the time-domain thermoreflectance (TDTR). In some previous simulations, Polanco et al [45] calculated the GaN-AlN interfacial thermal conductance to be 300 MW/(m$^2$K) using a nonequilibrium Green's function; Bao et al [46] predicted the interfacial thermal conductance of GaN-AlN by Monte Carlo method and NEMD simulation based on LJ potential, with calculated values of 870 MW/(m2K) and 780 MW/(m2K); Sun et al [47] investigated the GaN-AlN interfacial thermal conductance using three empirical interatomic potentials (SW potential, Tersoff-mixing rule, Transf. Tersoff), with final results of 937 MW/(m$^2$K), 3124 MW/(m$^2$K) and 3225 MW/(m$^2$K), respectively. It can be found that the prediction accuracy of our constructed NEP model is much higher than that of the traditional empirical potential and other numerical calculation methods, and it can be used to study the mechanism of the transition layer material and thickness on the heat transfer at the GaN-AlN interface.

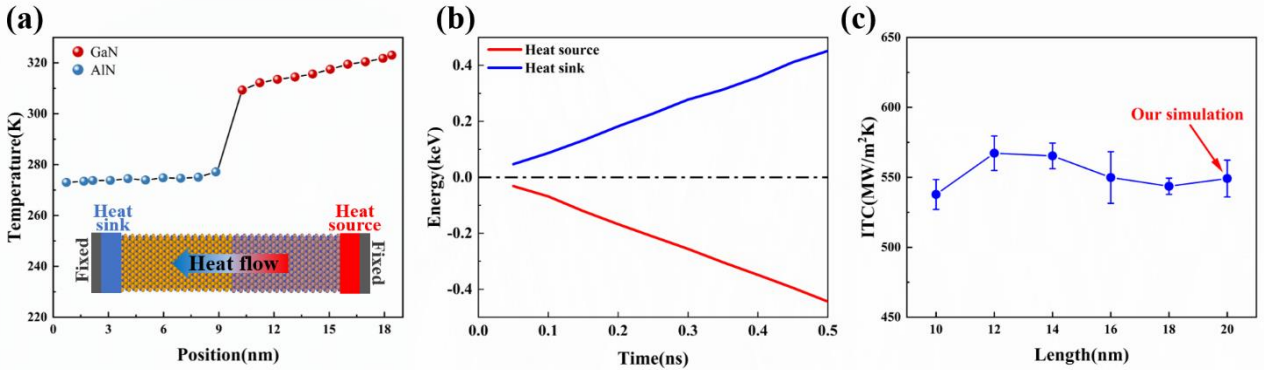

**Fig. 4.** (a) Temperature distribution in GaN-AlN heterostructure at steady state using NEP. (b) Variation of cumulative energy of heat source and heat sink with simulation time. (c) ITC values for different model lengths.

*3.2 Heat transfer at GaN-Al$_x$Ga$_{(1-x)}$N and AlN-Al$_x$Ga$_{(1-x)}$N heterostructure interfaces*

The NEMD simulations based on NEP further investigated the interfacial heat transfer of GaN-

$Al_xGa_{(1-x)}N$ (x=0.5-1) and $AlN$-$Al_xGa_{(1-x)}N$ (x=0-0.5) heterostructure interfaces at different Al component concentrations. A partial heterostructure model is shown in Fig. 5(a), where AlxGa(1-x)N is obtained by randomly replacing the corresponding proportion of Al(Ga) atoms inside the AlN(GaN) model with Ga(Al) atoms, and to be consistent with the real situation of the heterostructure interface of the GaN-based electronic device, the heat source of the GaN-$Al_xGa_{(1-x)}N$ structure is set on the GaN side, the heat sinks is set on the $Al_xGa_{(1-x)}N$ side, and the AlN-$Al_xGa_{(1-x)}N$ structure heat source is set on the $Al_xGa_{(1-x)}N$ side, and the heat sink is set on the AlN side. As shown in Fig. 5(b)(c), the thermal conductance of the GaN-$Al_xGa_{(1-x)}N$ (x=0.5-1) interface decreases with the increasing concentration of Al element in $Al_xGa_{(1-x)}N$, and the maximum value of ITC at x=0.5 is 1274 $MW/(m^2K)$, which is 128% higher compared with the GaN-AlN interface; whereas the thermal conductance of the AlN-$Al_xGa_{(1-x)}N$ (x=0-0.5) interface increases continuously with the increase of the Al elemental concentration, and the interfacial thermal conductance reaches the maximum value (ITC=1834 $MW/(m^2K)$) when x=0.5, which is a 229% increase in ITC compared to the GaN-AlN interface. It is noteworthy that although the ITC of the above two structures increases, the intrinsic thermal conductivity of the $Al_xGa_{(1-x)}N$ material obtained after atom substitution decreases significantly. As shown in Fig. 6(a)(b) for the temperature distribution of the heterostructure at 300 K with different Al elemental concentrations, the temperature drop at the interface is significantly reduced in the GaN-$Al_xGa_{(1-x)}N$ and AlN-$Al_xGa_{(1-x)}N$ structures compared to GaN-AlN, while the heat transferred from the heat source and heat sinks is also decreasing, whereas the thermal conductivity of the $Al_xGa_{(1-x)}N$ layer decreases continuously with increasing atomic doping ratio. For example, with GaN-$Al_xGa_{(1-x)}N$ heterostructure, when x=0.9 the doping ratio of Ga atoms is 0.1, the equivalent thermal resistance of the $Al_{0.9}Ga_{0.1}N$ layer is 1.32$m^2$K/GW, and when the doping ratio increases to 0.5 the equivalent thermal resistance of the $Al_{0.5}Ga_{0.5}N$ layer increases to 4.22$m^2$K/GW. This may be because elemental doping alters the otherwise stable crystal structure of the material, which significantly reduces the lattice thermal conductivity of the material, while the higher concentration of dopant atoms results in stronger phonon scattering in the region, leading to an increase in the local thermal resistance hindering the heat transport process. At this point, the intrinsic thermal resistance of the $Al_xGa_{(1-x)}N$ layer becomes the controlling thermal resistance in the heterostructure, and the temperature rise due to this thermal resistance is also much larger than that caused by the interfacial thermal resistance, such as in the GaN-$Al_{0.5}Ga_{0.5}N$ structure, where the temperature rise of the intrinsic thermal resistance of the $Al_{0.5}Ga_{0.5}N$ layer is 39 K, while the temperature rise resulting from the interfacial thermal resistance is only 10 K.

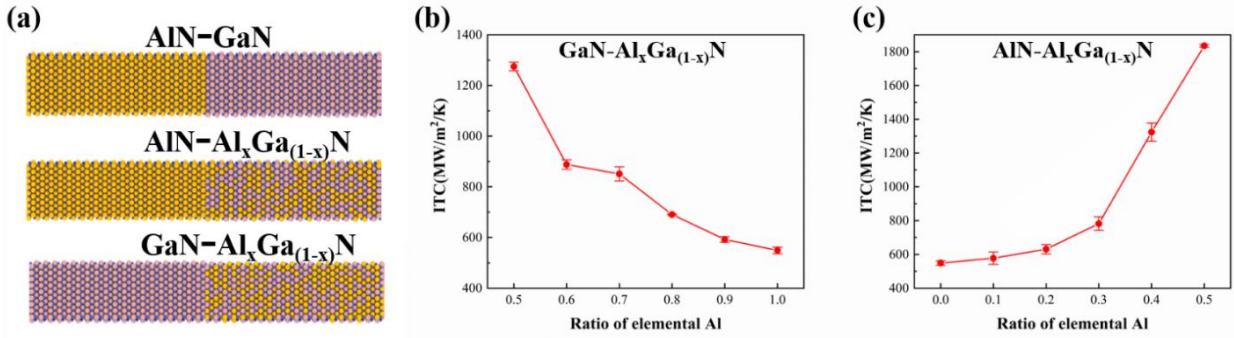

**Fig. 5.** (a) Partial AlN-GaN, AlN-Al$_x$Ga$_{(1-x)}$N, GaN-Al$_x$Ga$_{(1-x)}$N heterostructure models. Interface thermal conductivity of (b) GaN-Al$_x$Ga$_{(1-x)}$N and (c) AlN-Al$_x$Ga$_{(1-x)}$N at 300 K with different Al component concentrations.

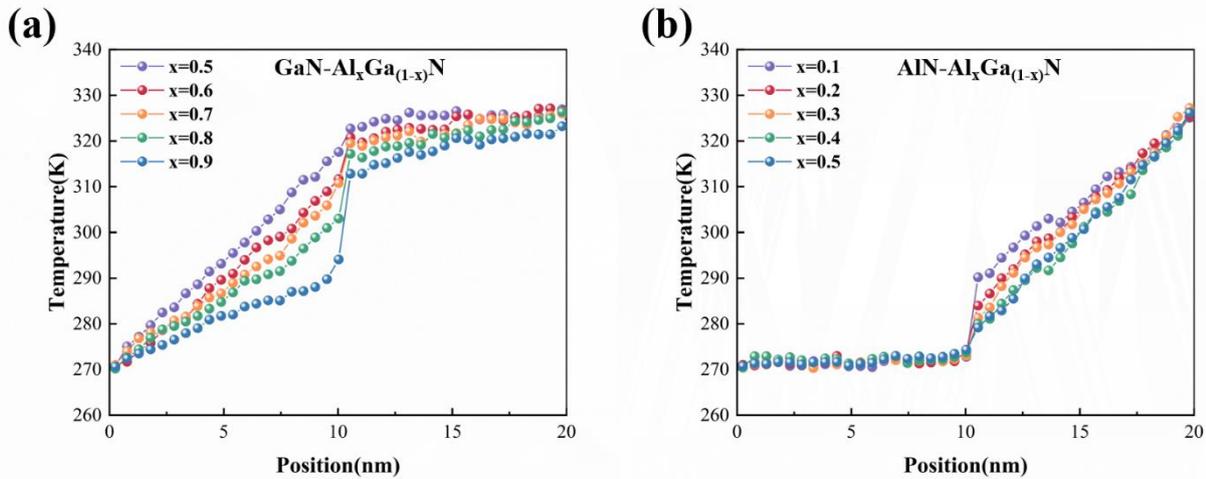

**Fig. 6.** Temperature distribution of heterostructure interfaces with different Al component concentrations at 300 K for (a) GaN-Al$_x$Ga$_{(1-x)}$N and (b) AlN-Al$_x$Ga$_{(1-x)}$N.

To investigate the heat transport mechanism at different interfaces and thus to analyze the reasons for the ITC changes, we performed phonon density of states (PDOS) calculations for the materials on both sides of the three heterostructure interfaces under study. As shown in Fig. 7(a) for the PDOS calculation results of the smooth GaN-AlN interface, the GaN phonons are mainly distributed in the frequency bands of 0-10 THz and 15-22 THz, and there is an obvious frequency gap in 10-15 THz, the AlN phonons are distributed in the frequency ranges of 0-25 THz, and there are two 7.6 THz and 19.3 THz PDOS peaks. The results of PDOS calculations at the GaN-Al$_{0.5}$Ga$_{0.5}$N interface are shown in Fig. 7(b). Compared with Fig. 7(a), the change of PDOS on the GaN side is not obvious, while the PDOS on the Al$_{0.5}$Ga$_{0.5}$N side changes significantly compared to that of the AlN, with the original low-frequency peak of 7.6 THz shifted leftward to the 5 THz, and the peak at the high-frequency of 19.3 THz is considerably reduced, resulting in an increased overlap area of the phonon density of states between the Al$_{0.5}$Ga$_{0.5}$N and GaN. Fig. 7(c) shows the PDOS distribution at the AlN-Al$_{0.5}$Ga$_{0.5}$N interface, the PDOS changes on the AlN side are very small compared to that at the GaN-AlN interface, whereas on the Al$_{0.5}$Ga$_{0.5}$N side the phonon scattering near the interface is enhanced due to the elemental doping thus promoting the phonon mode energy redistribution, $Al_{0.5}Ga_{0.5}N$ not only excited medium-frequency phonons in the 10-15 THz range compared to GaN, but also made the PDOS of $Al_{0.5}Ga_{0.5}N$ at the interface essentially overlap with AlN in the entire frequency interval. Similar conclusions such as the redistribution of phonon mode transport energy at the interface through light atom [48] or isotope [49] scattering, thus enhancing heat transfer have been investigated in previous articles. The increase of the phonon state density overlap region causes more elastic phonon scattering to occur at the interface, which promotes a more efficient transfer of interfacial heat flux, resulting in an increase of ITC at the $GaN-Al_xGa_{(1-x)}N$ and $AlN-Al_xGa_{(1-x)}N$ heterostructure interfaces.

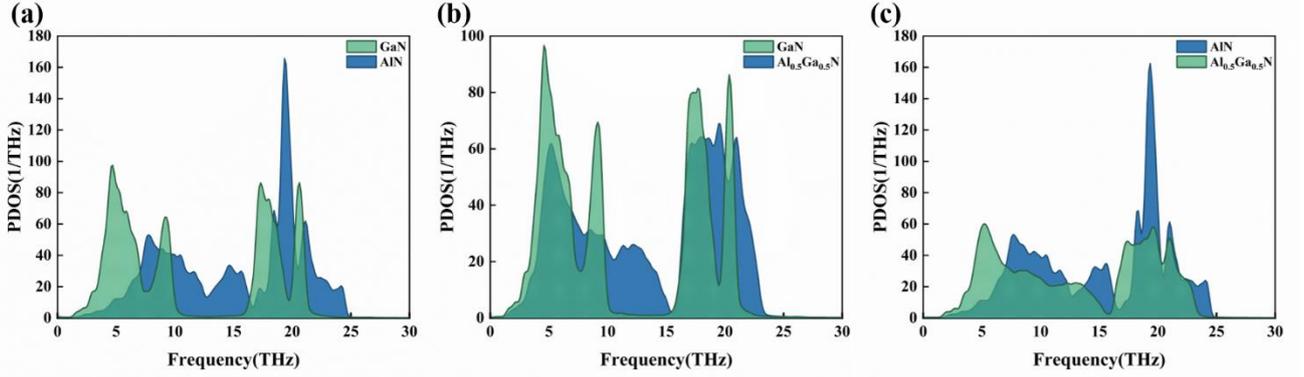

**Fig. 7.** The phonon density of states for materials on both sides of the heterostructure interface (a) GaN-AlN (b) $GaN-Al_{0.5}Ga_{0.5}N$ (c) $AlN-Al_{0.5}Ga_{0.5}N$.

To better understand the heat transfer mechanism at the interface, we performed spectral heat current (SHC) calculations for three areas of the GaN-AlN, $GaN-Al_{0.5}Ga_{0.5}N$, and $AlN-Al_{0.5}Ga_{0.5}N$ heterostructure interfaces, respectively, with areas A and B being the materials on both sides of the heterostructure, and area C is the area of the interface that contains the materials on both sides. For convenience of discussion, we divide the entire frequency range into three bands: low frequency (0-10 THz), medium frequency (10-15 THz) and high frequency (15-30 THz) sections. The results of the SHC calculations are shown in Fig. 8(a-f), it can be observed that all three heterostructure interfaces have a distinct SHC peak in both the low and high-frequency parts, which is expected because the PDOS of the materials on both sides of the three heterostructures overlap well near this frequency range. Meanwhile, the $GaN-Al_{0.5}Ga_{0.5}N$ and $AlN-Al_{0.5}Ga_{0.5}N$ interfaces show a significant enhancement of SHCs in the low and high-frequency regions compared to GaN-AlN, which leads to an increase in ITC. For the $GaN-Al_{0.5}Ga_{0.5}N$ structure, the phonon transport heat current is maximum at 5 THz, which corresponds well with the high overlap of the PDOS at 5 THz for the two interface materials in Fig. 7(b). In the $AlN-Al_{0.5}Ga_{0.5}N$ structure, the $Al_{0.5}Ga_{0.5}N$ side excites the medium-frequency phonons due to phonon scattering (Fig. 7(c)), and thus heat current can be observed to pass through the medium-frequency region of the SHC of the $AlN-Al_{0.5}Ga_{0.5}N$. A good match between the AlN and the $Al_{0.5}Ga_{0.5}N$ phonon spectra also results in the increased heat flux throughout the frequency range of the of transport. It can be found that the PDOS calculations can well explain the differences in the contribution of phonons of different frequencies to the heat current between heterostructure interfaces, while the consistency between PDOS and SHC also proves the accuracy of our NEP model calculations.

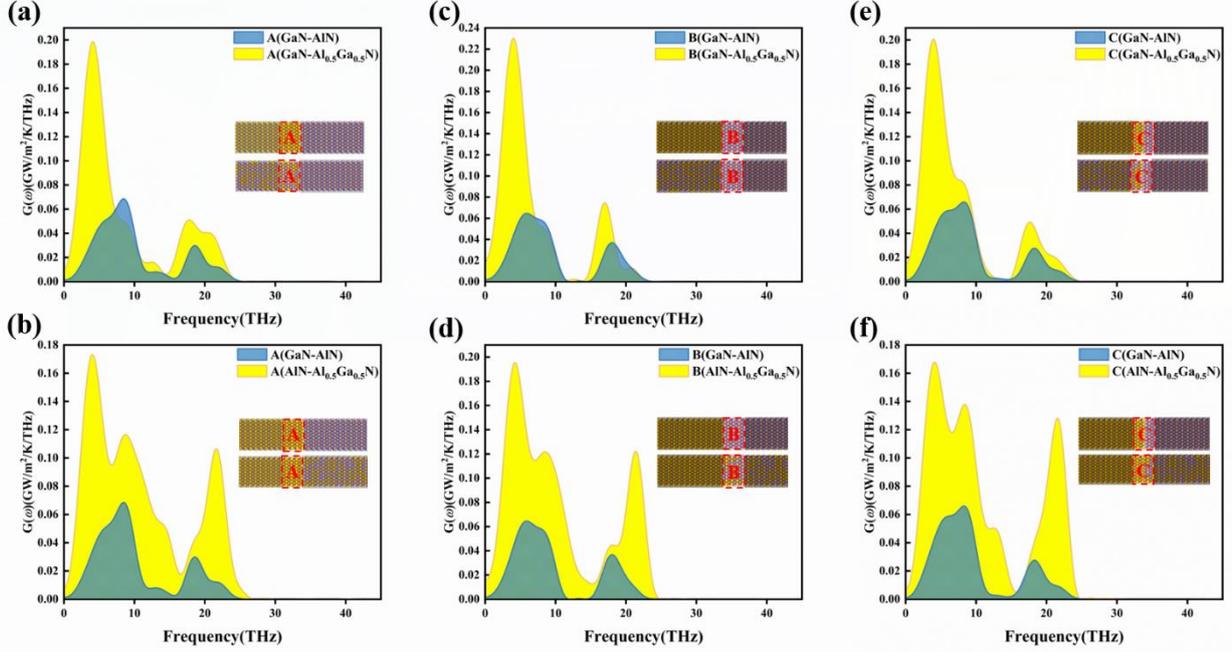

**Fig. 8.** SHC of GaN-AlN, GaN-Al$_{0.5}$Ga$_{0.5}$N and AlN-Al$_{0.5}$Ga$_{0.5}$N heterostructure interfaces in different regions (a-b) SHC of heterostructure interface region A, (c-d) SHC of heterostructure interface region B, (e-f) SHC of heterostructure interface region C.

*3.3 Effect of transition layer material and thickness on heat transfer at GaN-AlN interface*

Based on the above computational analyses of heat transfer at GaN-Al$_x$Ga$_{(1-x)}$N and AlN-Al$_x$Ga$_{(1-x)}$N heterostructure interfaces, the effects of different thicknesses (L$_{middle}$=0.5,1,1.5,2.5,5 nm) and compositions of Al$_x$Ga$_{(1-x)}$N (x=0.1-0.9) transition layers on the heat transfer at the GaN-AlN interface have been further investigated. A schematic of the GaN-Al$_{0.5}$Ga$_{0.5}$N-AlN structure is given as shown in Fig. 9(a), when the ITC at the GaN-AlN interface is the sum of the interfacial thermal resistance of Al$_x$Ga$_{(1-x)}$N with GaN and AlN and the bulk thermal resistance of the Al$_x$Ga$_{(1-x)}$N layer, which can be expressed as:

$$ITR_{GaN-AlN} = ITR_{GaN-Al_xGa_{(1-x)}N} + ITR_{AlN-Al_xGa_{(1-x)}N} + ITR_{Al_xGa_{(1-x)}N} (ITC_{GaN-AlN} = 1/ITR_{GaN-AlN}) \quad (6)$$

As shown in Fig. 9(b-f) are the calculation results of the ITC of GaN- Al$_x$Ga$_{(1-x)}$N-AlN interface in the interval range of transition layer thickness L$_{middle}$=0.5-5 nm, and the red dashed line is the ITC value of smooth GaN-AlN interface. When L$_{middle}$ is greater than or equal to 1.5 nm, the ITCs of GaN-Al$_x$Ga$_{(1-x)}$N-AlN interfaces containing the transition layer are all smaller than those of GaN-AlN interfaces without the transition layer. When L$_{middle}$=0.5 nm (Fig. 9(b)), with the increase of the Al elemental percentage in the AlxGa(1-x)N transition layer, the ITC first decreases (x=0-0.1) and then increases (0.1-0.7) before decreasing (x=0.7-0.9), at this time, the ITC of GaN-$Al_xGa_{(1-x)}$N-AlN structure containing transition layer with 20%-90% of Al element is larger than that of GaN-AlN structure, and ITC=711 MW/(m$^2$K) reaches the maximum when x=0.7, compared with the GaN-AlN structure without transition layer, the ITC is improved by 27.6%. When $L_{middle}$ = 1 nm (Fig. 9(c)), the $Al_xGa_{(1-x)}$N transition layer with Al elemental share greater than or equal to 60% can strengthen the GaN-AlN interfacial heat transfer, and the interfacial thermal conductance ITC = 608 MW/(m$^2$K) reaches the maximum value at x = 0.8. Thus there exists an optimum thickness and component concentration of $Al_xGa_{(1-x)}$N transition layer to enhance heat transfer at the GaN-AlN interface. As our results discussed in Section 3.2, the $Al_xGa_{(1-x)}$N transition layer can bridge the phonon spectra of the GaN and AlN layers to enhance the interfacial heat transfer, but as the atomic doping concentration increases, the bulk thermal resistance of the $Al_xGa_{(1-x)}$N layer rapidly increases to become a controlling resistance that will weaken the interfacial heat transfer, and thus the change of the interfacial thermal conductance will be affected by these two factors at the same time. As shown in Fig. 10 is a comparison of the spectral heat flux of the GaN-$Al_{0.7}Ga_{0.3}$N-AlN interface containing a 0.5 nm transition layer with that of the GaN-AlN interface without the transition layer, 0.5 nm $Al_{0.7}Ga_{0.3}$N interlayer bridges the phonon mismatch between GaN and AlN, which significantly increases the heat transferred by phonons in the frequency range of 0-10 THz, although the thermal conductivity of the $Al_{0.7}Ga_{0.3}$N interlayer is reduced, the combined effect is still to enhance the interfacial heat transfer. It is worth noting that, for heterostructure containing transition layers, there are differences in the rate of change of bulk thermal resistance of transition layers with different thicknesses and compositions, which can greatly affect the degree of reinforcement of thermal conductivity at the interface, so there are different optimal thicknesses and component concentrations for transition layers in different heterostructures.

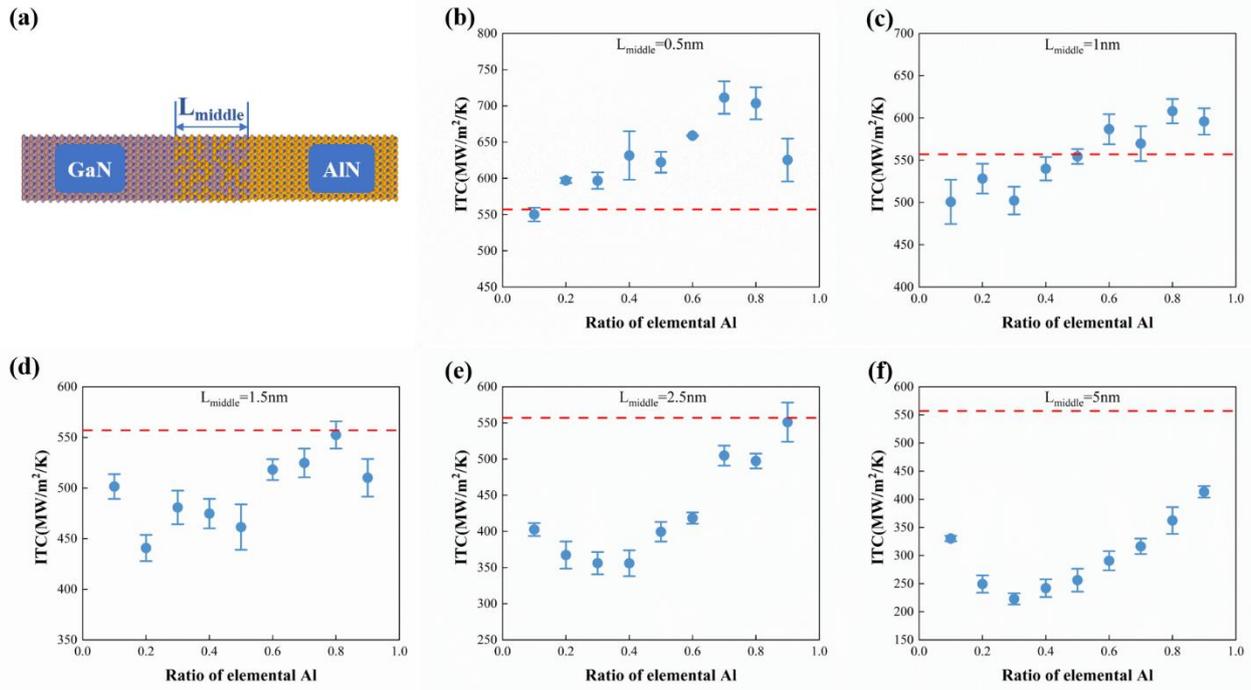

**Fig. 9.** (a) GaN-Al$_{0.5}$Ga$_{0.5}$N-AlN heterostructure model. (b-f) GaN-Al$_x$Ga$_{(1-x)}$N-AlN interfacial thermal conductance for different thicknesses and component concentration transition layers.

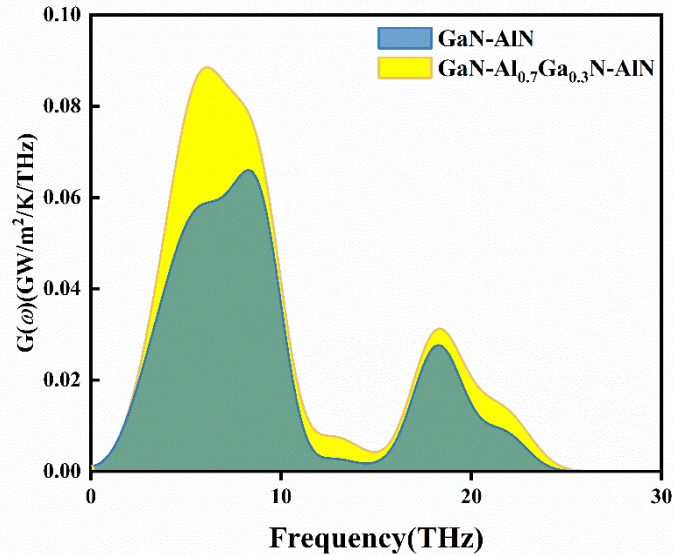

**Fig. 10.** Comparison of spectral heat current of GaN-Al$_{0.7}$Ga$_{0.3}$N-AlN interfaces containing a 0.5 nm transition layer and GaN-AlN interfaces without a transition layer.

## 4 Conclusion

A neural network-based machine learning potential NEP was trained by GPUMD for predicting GaN-AlN heterostructure interface heat transfer. Using NEP-based nonequilibrium molecular dynamics to calculate the ITC of GaN-AlN, GaN-Al$_x$Ga$_{(1-x)}$N, and AlN-Al$_x$Ga$_{(1-x)}$N heterostructure interfaces, and analyzing the interfacial heat transfer mechanism by PDOS and SHC, the following conclusions were obtained:

(1) When the Al element in $Al_xGa_{(1-x)}N$ is 50% (x=0.5), the ITCs of GaN-$Al_{0.5}Ga_{0.5}$N and AlN-$Al_{0.5}Ga_{0.5}$N are increased by 128% and 229% compared to GaN-AlN, which is attributed to the enhancement of phonon scattering near the interfaces due to the elemental doping and thus promotes the redistribution of the phonon energies, increasing the overlap area of $Al_xGa_{(1-x)}N$ with the PDOS of GaN and AlN, and more elastic phonon scattering occurs at the interface.

(2) For the 0.5 nm $Al_{0.7}Ga_{0.3}$N transition layer, the ITC of GaN-$Al_{0.7}Ga_{0.3}$N-AlN is improved by 27.6% compared to GaN-AlN. The atomically doped $Al_xGa_{(1-x)}N$ transition layer can bridge the phonon spectra of GaN and AlN to strengthen the interfacial heat transfer, but the volume thermal resistance of the $Al_xGa_{(1-x)}N$ layer increases rapidly with the increase of the doping ratio to become the control thermal resistance to weaken the interfacial heat transfer, so there exists an optimal thickness and concentration of Al components of the $Al_xGa_{(1-x)}N$ transition layer to strengthen the GaN-AlN interfacial heat transfer.

The findings of this work elucidate the mechanism of the influence of the material composition and thickness of the $Al_xGa_{(1-x)}N$ transition layer on the heat transfer at the GaN-AlN contact interface, providing theoretical insights into the efficient thermal management of heterostructure interfaces in GaN-based power devices.

**Acknowledgments**

This work was supported by the National Natural Science Foundation of China (Grant NO. U20A20299, 52006102), the Fundamental Research Funds for the Central Universities (No.30923010917).